\documentclass[prl,aps,a4paper,showpacs,twocolumn,floatfix]{revtex4}

\usepackage{graphicx}
\usepackage[ansinew]{inputenc}
\usepackage{array}
\usepackage{color}
\usepackage{amsmath}
\usepackage{amsxtra}
\usepackage{amstext}
\usepackage{amssymb}
\usepackage{latexsym}
\usepackage{dsfont}
\begin{document}

\title{Entangling the ro-vibrational modes of a macroscopic mirror using radiation pressure}

\author{M. Bhattacharya, P.-L. Giscard, and P. Meystre}
\affiliation{B2 Institute, Department of Physics and College of Optical
Sciences\\The University of Arizona, Tucson, Arizona 85721}

\date{\today}

\begin{abstract}
We consider the dynamics of a vibrating and rotating end-mirror of
an optical Fabry-P{\'erot} cavity that can sustain
Laguerre-Gaussian modes. We demonstrate theoretically that since
the intra-cavity field carries linear as well as angular
momentum, radiation pressure can create bipartite entanglement
between a vibrational and a rotational mode of the mirror. Further
we show that the ratio of vibrational and rotational couplings
with the radiation field can easily be adjusted experimentally,
which makes the generation and detection of entanglement robust to
uncertainties in the cavity manufacture. This constitutes the
first proposal to demonstrate entanglement between two
qualitatively different degrees of freedom of the same macroscopic
object.
\end{abstract}


\maketitle No known law of physics prevents the application of
quantum mechanics to macroscopic bodies. Characteristic traits of
quantum mechanics such as entanglement \cite{schrodinger1935} can
then in principle be displayed even by large objects. The
demonstration of quantum entanglement in macroscopic objects would
clearly be interesting from the point of view of fundamental
considerations, such as the exploration of the quantum-classical
boundary \cite{leggett2002}. It is also expected to have important
applied consequences since entanglement is a crucial resource for
information processing enabling quantum communication,
computation, and measurement, see Ref.~\cite{Nielsenbook} and
references therein.

Various mechanisms have been proposed for generating entanglement
between different degrees of freedom of macroscopic objects. The
flexural modes of nanomechanical electrodes can be entangled by
the ions they trap, via Coulomb interactions \cite{tian2004}.
Entanglement can be generated between the vibrations of an array
of gold beams fabricated on a semiconductor membrane using
electric voltages \cite{eisert2004}. The motion of a
nanomechanical oscillator carrying a ferromagnetic domain can
become entangled with the collective spin of a mesoscopic Bose
Einstein condensate due to their magnetic coupling
\cite{hansch2007note}. Radiation pressure can entangle two
vibrating nanofabricated mirrors that may belong to an optical
cavity
\cite{mancini2002,mancini2003,braunstein2003,pinard2005,tombesi2007}
or not \cite{pirandola2006}. In addition a single cavity mode can
also entangle multiple vibrational modes of the same mirror
\cite{mancini2003}.

In this Letter we discuss instead how radiation pressure can
entangle two \textit{qualitatively} different motional degrees of
freedom of the same classical object. Specifically we show that
the optomechanical coupling produced by a Laguerre-Gaussian intracavity
field can lead to bipartite entanglement between the modes of
rotation and vibration of a moving mirror. Further, we demonstrate
that the ratio of the coupling of the optical field to the
vibration and rotation modes can be adjusted
experimentally, resulting in the robust generation of entanglement
against uncertainties in the mass, radius and mechanical
frequencies of the mirror. This is in contrast to previous
proposals in which entanglement relies crucially on the precise
balance of mirror parameters, a situation that is difficult to
attain in practice \cite{mancini2002,mancini2003,vitali2003,
braunstein2003,pinard2005,tombesi2007}.

\begin{figure}[t]
\includegraphics[width=0.48 \textwidth]{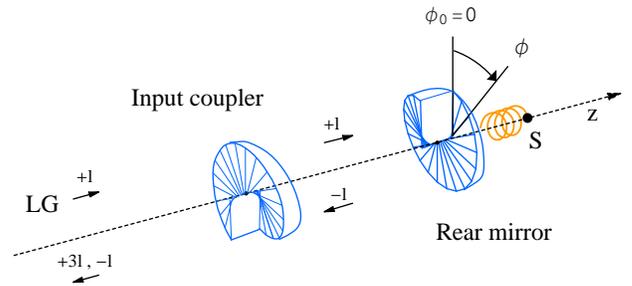}
\caption{\label{fig:cavitypic1}(Color online). The
arrangement proposed in this work for entangling the
vibrational and rotational modes of a mirror. A
Laguerre-Gaussian beam is incident on the resonant
cavity formed by two spiral phase elements, the
transmissive but fixed input coupler and the perfectly
reflective moving rear mirror. The rear mirror is mounted
on a helical spring $S$ which provides a vibrating restoring force
along the $z$ axis as well as torsion opposite the direction
$\phi$. The deflection of the rear mirror from its angular
equilibrium position $(\phi_{0}=0)$ is indicated by the angle
$\phi$; the $z$ deflection of the mirror has not been shown
for clarity. The charge on the Laguerre-Gaussian beams at
various points has been indicated.}
\end{figure}

The configuration that we consider consists of an optical cavity
formed by two spiral phase elements, see
Fig.~\ref{fig:cavitypic1}. Spiral phase elements can be reflective
or transmissive and are used to change the angular momentum or
`optical charge' of laser beams \cite{mishpm2}. The input coupler
transmits light weakly, and without changing its charge. A beam
reflected from it however gains a charge $2l$. The rear mirror on
the other hand reflects light perfectly, removing at the same time
a charge $2l$ from it. Designed in this manner the cavity can
provide mode build-up to an incident Laguerre-Gaussian field of
charge $+l$; a detailed discussion of the cavity conditions has
been given in Ref.~\cite{mishpm2}.

The input coupler is supported rigidly, but the rear mirror is
mounted in such a way that it can vibrate as well as rotate. One
way to accomplish this may be by mounting the mirror on a helical
spring \cite{kopf1990}. A number of experiments have demonstrated
that the linear vibrations of the mirror can be cooled by a
Gaussian cavity mode \cite{mirrorcool}; in a recent proposal we
have shown that if a Laguerre-Gaussian mode is used instead
rotational cooling can also be achieved \cite{mishpm2}. In the
current design we combine both effects -- since each
Laguerre-Gaussian photon carries linear as well as angular
momentum, the same cavity mode can affect both the vibration as
well as the rotation of the mirror.

The coupling of radiation to the mirror motion can be derived by
using the fact that both the linear and the angular momentum of
the incident $+l$ Laguerre-Gaussian beam are reversed by the rear
mirror. The torque per intracavity photon is the rate of
change of angular momentum $2l\hbar/(2L/c)$, and similarly for the
force, with rate of change of momentum $2\hbar k/(2L/c)$. Here $k$
is the wave vector of the light field, $L$ is the length of the
cavity and $c$ is the velocity of light \cite{mishpm2}.

Using these arguments we can model the physical system described
above and shown in Fig.~\ref{fig:cavitypic1} by the Hamiltonian
\begin{eqnarray}
\label{eq:Ham} H &=& \hbar
\omega_{c}a^{\dagger}a+\frac{\hbar\omega_{z}}{2}(p_{z}^{2}+z^{2})
+\frac{\hbar\omega_{\phi}}{2}(L_{z}^{2}+\phi^{2}) \nonumber
\\
&-&\hbar g_{z} a^{\dagger} a z+\hbar g_{\phi} a^{\dagger} a \phi.
\end{eqnarray}
The first term in this Hamiltonian describes the electromagnetic 
energy of the cavity mode, the next two terms the vibrational and 
rotational energies of the moving mirror and the last two terms 
the effects of radiation force and torque on the rear mirror 
respectively. 

In Eq.~(\ref{eq:Ham}) $a$ and $a^\dagger$ are the 
bosonic annihilation and creation operators for the cavity mode.
$z$ and $p_{z}$ are the dimensionless position and momentum of 
the mirror scaled to the characteristic length $(\hbar/M\omega_z)^{1/2}$
and momentum $(\hbar M\omega_z)^{1/2}$, $M$ being the mass of the 
mirror and $\omega_z$ its frequency of linear vibration. 
Similarly $\phi$ and $L_{z}$ are the dimensionless mirror 
angular displacement and momentum scaled to $(\hbar/I\omega_\phi)^{1/2}$ 
and $(\hbar I\omega_\phi)^{1/2}$ respectively, $I=MR^{2}/2$ being 
the moment of inertia, $R$ the mirror radius, and $\omega_\phi$ the 
frequency of angular vibration. The commutation 
relations of the dynamical variables are given by 
$[a,a^\dagger]=1,[z,p_{z}]=i$, and 
$[\phi,L_{z}]=i$ respectively. The frequency 
$\omega_{c}=n\pi c/L$ is the cavity mode frequency, 
where $L = n\lambda/2$. Finally, the \textit{opto-vibrational} and 
\textit{opto-rotational }coupling constants are given by
\begin{equation}
 \label{eq:couple}
g_{z}=\frac{\omega_{c}}{L}\sqrt{\frac{\hbar}{M\omega_{z}}},\,\, \,\,\,
g_{\phi}=\frac{cl}{L}\sqrt{\frac{\hbar}{I\omega_{\phi}}},
\end{equation}
respectively.

We consider the Heisenberg equations of motion for the dynamical
variables of the Hamiltonian (\ref{eq:Ham}), adding damping and
noise to arrive at the nonlinear quantum Langevin equations for
the system \cite{gardinerbook}:
\begin{eqnarray}
\label{eq:QLErot}
\dot{a}&=& -i(\delta-g_{z}z+g_{\phi}\phi)a-\frac{\gamma}{2}a+\sqrt{\gamma}a^{\rm in},\nonumber \\
\dot{z}&=&\omega_{z}p_{z}, \nonumber \\
\dot{p_{z}}&=&-\omega_{z}z+g_{z}a^{\dagger}a-\gamma_{z} p_{z}+\epsilon_{z}^{\rm in}, \\
\dot{\phi}&=& \omega_{\phi}L_{z},\nonumber \\
\dot{L_{z}}&=&  -\omega_{\phi}\phi-g_{\phi}a^{\dagger}a
-\gamma_{\phi}L_{z}+\epsilon_{\phi}^{\rm in}.\nonumber \\
\nonumber
\end{eqnarray}
Here $\delta =\omega_{c}-\omega_{L}$ is the detuning of the laser
frequency $\omega_{L}$ from the cavity resonance, $\gamma$ is the
damping rate of the cavity, $\gamma_z$ and $\gamma_\phi$ the
intrinsic damping rates of vibration and rotation respectively, 
and $a^{\rm in}$ is a noise operator describing the laser field 
incident on the cavity. The mean value $\langle a^{\rm in}(t)\rangle =a_{s}^{\rm
in}$ describes the classical Laguerre-Gaussian field, and the
delta-correlated fluctuations
\begin{equation}
\label{eq:fluc}
\langle \delta a^{\rm in}(t) \delta a^{\rm in,\dagger}(t')
\rangle=\delta (t-t'),
\end{equation}
describe the vacuum noise injected into the cavity mode by the
driving field. The Brownian noise operator $\epsilon_z^{\rm in}$
accounts for the mechanical noise that couples into the mode of
mirror vibration from the thermal environment. Its mean value is
zero and its fluctuations are correlated at temperature $T$ as
\cite{gardinerbook}
\begin{eqnarray}
\label{eq:Brownian} &&\langle \delta \epsilon_z^{\rm in}(t) \delta
\epsilon_z^{\rm in}(t') \rangle=\nonumber \\
&&\frac{\gamma_z}{\omega_z} \int_{-\infty}^{\infty}
\frac{d\omega}{2\pi}e^{-i\omega(t-t')}\omega
\left[1+ \coth \left(\frac{\hbar \omega}{2k_{B}T}\right)\right],\nonumber \\
\end{eqnarray}
where $k_{B}$ is Boltzmann's constant. Similar relations hold for
the rotational noise operator $\epsilon_\phi^{\rm in}$.

The steady-state values of the dynamical variables can be found
from the equations
\begin{eqnarray}
\label{eq:sstate}
a_{s}&=&\frac{\sqrt{\gamma}|a_{s}^{\rm in}|}{\left[(\frac{\gamma}{2})^{2}+(\delta-a_{s}^{2}G)^{2}\right]^{1/2}}, \nonumber \\
 z_{s}&=&\frac{g_{z}a_{s}^{2}}{\omega_{z}}, \,\,\,\,\,\phi_{s}=-\frac{g_{\phi}a_{s}^{2}}{\omega_{\phi}}, \\
 p_{z,s}&=&0, \,\,\,\,\,\,L_{z,s}=0, \nonumber \\ \nonumber
\end{eqnarray}
where $G=g_z^2/\omega_z+ g_\phi^2/\omega_\phi$, and the phase of
the input field $a_{s}^{\rm in}$ has been chosen such that $a_{s}$
is real. The field amplitude $a_{s}$ is found by solving the first
equation, which is nonlinear, and $z_{s}$ and $\phi_{s}$ can then
be determined. The solutions to Eq.~(\ref{eq:sstate}) display
bistability for high enough input power $P_{\rm in}=\hbar
\omega_{c}|a_{s}^{\rm in}|^{2}$ \cite{pm1985}. In the rest of the
paper we assume the use of electronic feedback, which allows us to
set the net detuning $\Delta =\delta-a_{s}^{2}G$ independently of
radiation pressure and also to suppress bistability. Such a
procedure is carried out routinely in mirror cooling experiments
\cite{mirrorcool}.

To investigate the behavior of the system for small deviations
away from its steady-state we expand every operator as the sum of
a (steady state) mean value [Eq.~(\ref{eq:sstate})] and a small
fluctuation, e.g. $a=a_{s}+\delta a$. Treating the other operators
in Eq.~(\ref{eq:QLErot}) in a similar way and retaining only terms
linear in the fluctuations yields
\begin{eqnarray}
\label{eq:fluctrot}
\dot{\delta a}&=& -(i\Delta+\frac{\gamma}{2})\delta a+i a_{s}(g_{z}\delta z-g_{\phi}\delta\phi) +\sqrt{\gamma}\delta a^{\rm in}, \nonumber \\
\dot{\delta z}&=&\omega_{z}\delta p_{z}, \nonumber \\
\dot{\delta p_{z}}&=&-\omega_{z}\delta z+g_{z}a_{s}(\delta a+\delta a^{\dagger})-\gamma_{z} \delta p_{z}+\delta \epsilon_{z}^{\rm in},\\
\dot{\delta \phi}&=&\omega_{\phi}\delta L_{z},\nonumber \\
\dot{\delta L_{z}}&=&-\omega_{\phi}\delta\phi-g_{\phi}a_{s}(\delta
a+\delta a^{\dagger}) -\gamma_{\phi}\delta
L_{z}+\delta\epsilon_{\phi}^{\rm in}.\nonumber \\  \nonumber
\end{eqnarray}
We solve Eq.~(\ref{eq:fluctrot}) in the frequency domain
\cite{vitali2007}. Combining the solutions of
Eq.~(\ref{eq:fluctrot}) with the relations in Eqs.~(\ref{eq:fluc})
and (\ref{eq:Brownian}) allows one to obtain the correlations
between the quantum fluctuations of the dynamical variables, and
hence the entanglement in the system \cite{mancini2002}. This is
because the fluctuations are continuous Gaussian variables fully
determined by their first and second moments. Computable measures
for bipartite entanglement between such variables exist and have
been used previously to quantify optomechanical systems
\cite{vitali2007}. The calculation of entanglement in the
frequency domain is also appropriate since the cavity dynamics are
experimentally easier to probe spectrally than in the time domain
\cite{mancini2003}.

More specifically, we consider the operators
$\delta u=\delta z-\delta \phi$ and $\delta v=\delta p_{z}+\delta L_{z}$.
We then construct the corresponding Hermitian
operators
$\mathcal{R}_{u,v}$, where
$\mathcal{R}_{u}=[\delta u (\omega)+\delta u (-\omega)]/2$,
for example. An entanglement measure $\mathcal{E}(\omega)$ can
then be defined as
\cite{mancini2002}
\begin{equation}
\label{eq:measureent}
\mathcal{E}(\omega)=\frac{\langle \mathcal{R}_{u}^{2}(\omega)\rangle \langle \mathcal{R}_{v}^{2}(\omega)\rangle }
{ |\langle [\mathcal{R}_{z}(\omega),\mathcal{R}_{p_{z}}(\omega)]\rangle |^{2} }.
\end{equation}
Ro-vibrational entanglement exists at the system response
frequency $\omega$ whenever  $\mathcal{E}(\omega)<1$.

\begin{figure}[t]
\includegraphics[width=0.50 \textwidth]{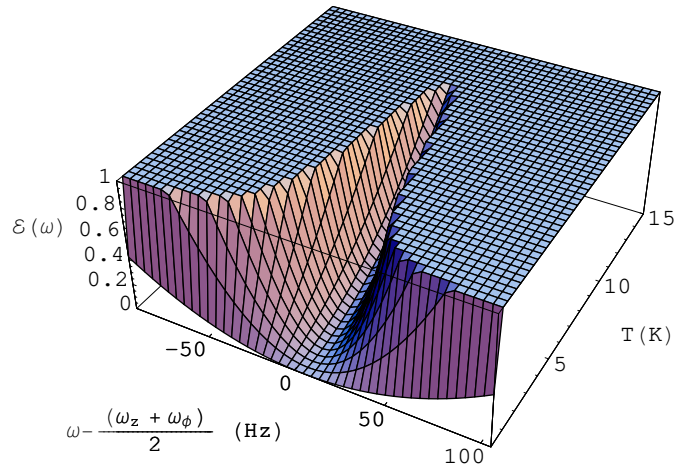}
\caption{\label{fig:cavitypic2}(Color online). Radiation
pressure-induced entanglement $\mathcal{E}(\omega)$ between a
vibrational and a rotational mode of the moving mirror. The
parameters are $M=1\mu$g, $R=15\mu$m, vibrational and rotational
quality factors $Q_{z}=Q_{\phi}=10^{6}$, $\omega_{z} \simeq
\omega_{\phi} =1$ MHz, $l=82$, $L \simeq 4$ mm, cavity finesse
$F=2.5\times 10^{4}$, $\lambda =812.7$nm, $\Delta =
\omega_{\phi}$, and $P_{in}=1$mW.}
\end{figure}

Figure~\ref{fig:cavitypic2} plots $\mathcal{E}(\omega)$ as a
function of the response frequency $\omega$ and temperature $T$
for the experimentally accessible parameters detailed in the
caption. A significant amount of entanglement is available at
higher than cryogenic temperatures, a regime in which mirror
cooling has been demonstrated \cite{rugar2007}, and at a usably
large bandwidth.

\begin{figure}[b]
\includegraphics[width=0.48 \textwidth]{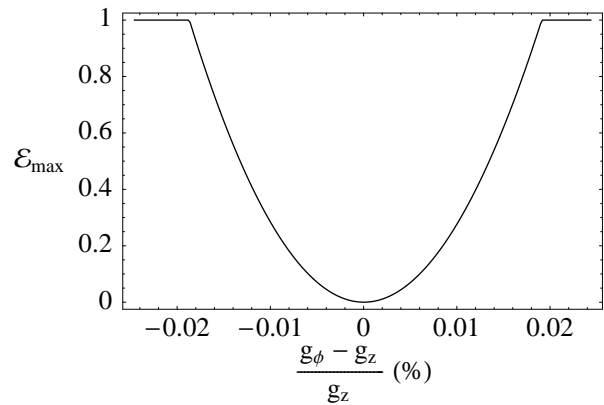}
\caption{\label{fig:cavitypic3} Maximum entanglement
$\mathcal{E}_{\rm max}$ at $T=1$K between a vibration and a
rotational mode of the rear mirror of Fig.~\ref{fig:cavitypic1} 
as a function of the percent fractional imbalance in the 
couplings with the radiation field.
The entanglement decreases with increasing asymmetry in the
couplings. In the figure for a 200 Hz difference between
rotational and vibrational frequencies centered at 1 MHz, the
coupling imbalance, about $0.02\%$, destroys the entanglement
completely. The remaining parameters are the same as in
Fig.~\ref{fig:cavitypic2}.}
\end{figure}

We observe that the entanglement is always maximum at the
arithmetic mean of the two mechanical frequencies, i.e.
$\mathcal{E}_{\rm max}=\mathcal{E}
\left((\omega_z+\omega_\phi)/2\right)$ \cite{mancini2002,mancini2003}.
Further, we find the presence of the  symmetry $g_{z}=g_{\phi}$ to
be crucial to the generation of entanglement in our system. This
is illustrated in Fig.~\ref{fig:cavitypic3}
which plots the maximum entanglement $\mathcal{E}_{\rm max}$ as a
function of the percent fractional imbalance in the couplings. It
shows that entanglement vanishes rapidly even for small deviations
away from equality. Further investigations indicate that for
increasing coupling asymmetry the bandwidth of entanglement also
vanishes faster with temperature. The requirement of a symmetric
coupling in order to radiation-entangle mechanical modes has been
noted previously in the case of mirror vibration
\cite{mancini2002,mancini2003,vitali2003,
braunstein2003,pinard2005,tombesi2007}. Very recently the effect
of asymmetry has been precisely characterized for the
case of general Gaussian continuous variables \cite{law2007}. It
has been shown analytically that higher the asymmetry, lower the
entanglement.

It can be seen from the expression (\ref{eq:couple}) for $g_{z}$
that for the case of entanglement between two vibrational modes of
the same mirror \cite{mancini2003} the ratio of the couplings is
determined by the frequencies of these modes. These may be
difficult to control experimentally and are typically unequal. For
our system the two mirror degrees of freedom couple
differently to the field, with a ratio
\begin{equation}
 \label{eq:coupleratio}
\frac{g_{z}}{g_{\phi}}=\frac{2\pi}{l\lambda}\sqrt{\frac{I\omega_{\phi}}{M\omega_{z}}}.
\end{equation}
In case the frequencies $\omega_z$ and $\omega_\phi$, mass $M$ and
moment of inertia $I$ are all slightly different from their
nominal values, it is possible to equalize the couplings by
varying the radiation wavelength $\lambda$, simultaneously
adjusting the cavity length $L$ so as to stay on resonance. 

We
have used this procedure to arrive at the values presented in
Fig.~\ref{fig:cavitypic2}, where the imbalance in the couplings is
assumed to be due to a frequency mismatch
$\omega_{\phi}-\omega_{z}=10$ Hz. We found that the couplings
could be equalized and the entanglement retained by tuning
$\lambda$ by $\sim 2$nm and the cavity length by $\sim 100\mu$m.
Such adjustments are easily within reach of current experimental
techniques. We finally note that the experimental measurement of
the entanglement $\mathcal{E}(\omega)$ can be carried out using
standard techniques such as homodyne measurements as described in
Ref.~\cite{mancini2002}. This involves use of a secondary cavity
with a third mirror beyond the rear mirror.

In conclusion we have demonstrated that a vibrational and a
rotational mode of the same macroscopic mirror can be entangled
quantum mechanically by radiation pressure from a
Laguerre-Gaussian cavity mode. The entanglement can be made robust
against imprecision in the cavity manufacture because the ratio of
vibrational-to-rotational coupling with the radiation can be tuned
experimentally.

This work is supported in part by the US Office of Naval Research,
by the National Science Foundation, and by the US Army Research
Office. We thank H. Uys and O. Dutta for helpful conversations.

\end{document}